\input harvmac
\rightline{FT98-6}
\Title{}{On the Initial Conditions for Pre-Big-Bang Cosmology}

\centerline{M. Borunda and M. Ruiz-Altaba{$^\ddagger$}
\footnote {}{$^\ddagger$ monica, marti @ft.ifisicacu.unam.mx}}
\bigskip\centerline{Departamento de F\'\i sica Te\'orica}
\centerline{Instituto de F\'{\i}sica}
\centerline{Universidad Nacional Aut\'onoma de M\'exico}
\centerline {Apartado Postal  20-364} \centerline{01000 M\'exico, D.F}

{\bf Abstract} :  The beautiful scenario of pre-big-bang cosmology is 
appealling not only because it is more or less derived from string theory, 
but 
also because it separates clearly the problem of the initial conditions 
for the 
universe from that of high curvatures. Recently, 
the pre-big-bang program was 
subject to attack from  on 
the grounds that pre-big-bang cosmology does not solve the horizon and flatness 
problems in a ``natural'' way, as customary exponential ``new'' inflation does. 
In particular, it appears that an arbitrarily small deviation from perfect 
flatness  in the initial state can not be accommodated.  For this analysis,  
matter in the universe before the big bang was assumed to be radiation. 
We perform a similar analysis to theirs, but using the equation of 
state for ``string matter'' $\rho=-3p$ which seems more appropriate to the 
physical situation and, also, is motivated by the scale factor duality 
(in the 
flat case) with respect to our expanding, radiation dominated, 
universe. For an open universe we find, exactly, the same 
time-dependence of the scale factor 
as in the Milne universe, recently found to represent the 
universal attractor 
at $t=-\infty $ of all pre big bang cosmologies. 
We conclude that our radiation dominated universe comes from a 
flat  rather than a curved region.

\newsec{Introduction}

 The
standard cosmological model does not deal with the initial 
singularity problem. In
order to solve this Veneziano and collaborators have 
developed a stringy
cosmology \ref\rone{M. Gasperini and 
G. Veneziano,{\it Astropart. Phys.} 1, 317
(1993)}\ref\ronee{   G. Veneziano, {\it Phys. Lett.} 
{\bf B}265, 287 (1991)}\ref\roneee{ M. Gasperini and G.
Veneziano, {\it Phys. Rev.} {\bf D}50, 2519 (1994)}
\ref\two{R. Brustein and G. Veneziano, {\it Phys. Lett.} {\bf B}329, 429 (1994)
} \ref\four{E.J.
Copeland, Amitabha Lahiri and D. Wands, 
{\it Phys. Rev.} {\bf D}50, 4868 (1994)}
\ref\reasther {R. Easther and K.
Maeda, {\it Phys. Rev.} {\bf D}54, 7252 (1996)},  mainly based on a
property of the low-energy effective action, called 
scale-factor-duality. It is now possible to think physically 
about the universe before the high curvature, high 
temperature regime called the big bang. Recently, the 
horizon problem in the pre-big bang scenario has attracted a 
great deal of attention
\ref\rthre{G. Veneziano, {\it Phys. Lett.} {\bf B}406, 297 (1997)}
\ref\rthreeee{  A. Buonanno, K.A. Meissner, C. Ungarelli
and G. Veneziano, {\it Phys. Rev.} {\bf D}57, 2543 (1998) }
 \ref\rthree {M. Turner and E. Weinberg {\it Phys.Rev.} {\bf D}56, 4604 (1997) }
\ref\rthreee{ N. Kaloper, A. Linde and R. Bousso, hep-th/9801073}
\ref\rthreeee{M. Maggiore and R. Sturani, {\it Phys. Lett.} 
{\bf B}415, 335 (1997)} assuming that the universe, 
before the big bang was radiation dominated. There is, 
however, no reason to expect that. Rather, within the 
validity range of the effective action for the massless 
modes of the strings, we expect all matter to still behave 
stringly, effectively as large null strings.

In section 2, motivated by the
scale-factor duality,  we use the equation of state 
of string matter  in a homogeneous universe. Section 3 shows how these
solutions deal with the horizon problem,  and we conclude in section 4.

\newsec{Cosmic solutions}
The equations describing the universe during the 
pre-big-bang phase ($t<0$) are obtained from the 
tree-level low-energy effective action of strings 
\ref\cuerdas{C. Vafa, hep-th/9702201}
\eqn\action{S=-{1\over 16\pi G}\int d^4x 
\sqrt{|g|}e^{-\phi }(R+\partial_\mu\phi\partial^\mu\phi ) 
+\sum_{\rm matter}\int d^4x\sqrt{|g|}{\cal L}_{\rm matter}}
where $g_{\mu\nu}$ is the four dimensional space-time metric, 
$\phi$ is the dilaton field, and ${\cal L}_{\rm matter}$ 
accounts for  all other fields (including the Kalb-Ramond field).

In  a Friedman-Robertson-Walker  background, the metric is
\eqn\metrica{ds^2=dt^2-a^2(t)\left\{ {dr^2\over 1-kr^2}+
r^2d\theta +r^2\sin ^2\theta d\phi^2\right\}}
where $a(t)$ is  the cosmic scale factor and $k=0,\pm 1$.  
The equation of motion for a homogeneous  dilaton $\phi (t) $ is 
\eqn\dila{\dot\phi^2 -2\ddot\phi -6\dot\phi H +
6{\ddot a\over a}+6H^2+6{k\over a^2}=0}
where $H=\dot a/a$ is the Hubble constant.

The Einstein equations are
\eqn\mate{2(R_\mu^\nu +\nabla_\mu\nabla^\nu\phi )=
16\pi Ge^\phi T_\mu^\nu}
where $T_{\mu\nu}$ is derived from ${\cal L}_{\rm matter}$.
By combining the time component of   \mate\ with  \dila, 
we get the equation for the energy density
\eqn\densidad{6H^2-6\dot\phi H +\dot\phi^2 + 6{k\over a^2}=
16\pi Ge^\phi\rho }
and from the space component of   \mate, the equation for the pressure
\eqn\presion {3H^2+\dot H -H\dot\phi -2{k\over a^2}=8\pi Ge^\phi p}

The three equations \dila\ ,\densidad\ and 
 \presion\  describe a homogenous and isotropic  
  universe (we disregard throughout spatial fluctuations, 
  otherwise important for structure formation). 
  In order to solve them we need an equation of state. 
  Using radiation $\rho =3p$, it is easy to see that one 
  solution is given by our post big bang ($t>0$) radiation 
  dominated universe, $k=0$, $a(t)\sim t^{1/2}$, $\phi =$constant. 
  Using the cosmological form of $T$-duality dubbed scale-factor 
  duality \rone\  this solution is mapped to a pre-big-bang 
  universe with increasing Newton's constant
\eqn\newton {G_N=\ell_{st}^2 e^{\phi}}
which satisfies equations \dila\ -\presion\ with ($t<0$),
\eqn\soltres{a(t)=a_0\left ( -{t\over t_0}\right )^{-1/2}}
\eqn\soltrese{ \phi (t)= \phi _0 -3\ln \left ( -{t\over t_0}\right )}
This universe, with initially vanishing Newton's 
constant and scale factor, i.e., initially flat and 
empty, is the reasonable initial condition for a 
stringy cosmology. As emphasized by Veneziano, 
this initial condition is no less reasonable than 
the usual one in standard inflationary cosmology, 
namely a very hot and curved one. The interest of 
the pre-big bang scenario is that the initial condition 
is a separate issue from the behavior of matter at high 
temperature and energy.
Interestingly enough, the duality transformation not only 
takes $t$ to $-t$ and $a$ to $1/a$, but it also changes the 
dilaton and, most curiously, changes also the equation of state
 \rone\ \eqn\estado {\rho =-3p}
This equation of state represents string matter 
\ref\rbowick{M. Bowick and L. Wijewardhana, 
{\it Phys. Rev. Lett.} 54, 2485, 1985}.

We wish to investigate the stability under a 
variety of initial conditions of the above pre-big-bang scenario. 
In particular, we are interested in whether enough inflation can
 take place while $t<0$ to solve the problems of standard 
 non-inflationary cosmology. Following the ideas in \rthree,  
 we  study what happens when $k\neq 0$. Note that, in this case, 
 there is no abelian T-duality, and the non-abelian duality  
 exists but  it  has no spherical symetry
  \ref\ragregada{X. Ossa and F. Quevedo, 
  {\it Nucl. Phys. }{\bf  B}403, 377 (1993)  }. 
  Thus, we cannot find pre-big-bang cosmological solutions 
  from post big bang ones, but of course we can just 
   solve the equations directly. Since the equation of 
   state describing stringy matter is a local feature of the 
   equations, it should be insensitive to the global nature 
   of the universe. We thus use, for $t<0$, the same equation 
   of state \estado\ for $k=\pm1$ as for $k=0$. At very early 
   times ($t\to -\infty$, $\phi\to -\infty$ ), 
all interactions are turned off, and thus what equation of 
state we use to 
simulate the dynamics of all the fields other than the metric 
and the dilaton is 
actually irrelevant. However, in  later stages of the 
pre-big-bang evolution, 
when gravity starts inflating the universe, it 
should make a difference. 
We are not interested in periodic universes, so 
we will concentrate on $k=0$ and $k=-1$.

For an open universe, with  spatial  curvature  $k=-1$, 
we find two possible universes
\eqn\soluno{\eqalign{
a(t)&={1\over\sqrt 3}(-t)\cr
\phi (t)&=\phi _0 + 6\ln \left (-{t\over t_0} \right )
}}
and
\eqn\soldos{\eqalign{a(t)&=\sqrt 3 (-t)\cr 
\phi (t) &=\phi _0 + 2\ln \left (-{t\over t_0}\right ) }}

As we can see, the dilaton field is  singular at 
both ends of the time evolution, $t\to -\infty$ and 
$t\to 0^-$. However, this does not mean that there 
is a strong-coupling regime in the far past. At very 
early times the  equations of motion are a bit different 
because spatial and time derivatives are of comparable 
importance and in this treatment we are neglecting spatial 
derivatives. As  shown in 
\ref\rmilne{A. Buonanno, K.A. Meissner, C. Ungarelli
 and G. Veneziano, {\it Phys. Rev.} {\bf D}57, 2543 (1998) }  
 once the spatial gradients are taken into account there 
 is a generic early-time attractor which corresponds to the Milne universe, which eventually (as $t$ goes from $-\infty$ to $0$) would turn  in ours solution.
 
To map the above solutions to the Einstein frame, 
we use the conformal 
rescaling   \ref\rdos{M. Gasperini and G. Veneziano, 
{\it Mod. Phys. Lett. A}{\bf  8}, 3701 (1993)}
\eqn\transformaciones {\eqalign{\tilde{g}_{\mu\nu}&=
g_{\mu\nu}e^{-\phi }\cr \tilde{\phi}&=\phi\cr}} 
whereby  the scale factor for $k=0$ takes the form
\eqn\einone {\tilde {a} (\tilde {t})\sim (-\tilde {t})^{2/5}}
 whereas the scale factor for the  solution \soluno\ with $k=-1$ goes like
\eqn\eintwo  {\tilde {a}(\tilde {t})\sim( -\tilde {t})}
while for  \soldos\  is constant and thus all the dynamics are governed by the dilaton.

Note that neither in the string frame nor in the  Einstein 
frame we do get an accelerated solution for $k=-1$ since
 $\ddot {a}(t)=\ddot {{\tilde a}}({\tilde t})=0$, 
 \ref\rolvidame {D. Clancy, J.E. Lidsey and R. Tavakol, gr-qc/9802052}. 
 This is of some relevance, because the pre-big-bang 
 flat universe is  accelerated corresponding to a post-big-bang 
 decelerated universe (ours), but for an open universe the 
 pre-big bang solution is linear in time. This is not very 
 appealling, as we shall see. Still, due to the ugly 
 behavior of the dilaton, equations \soluno\ and \soldos, 
 much care should be exercised to interpet the solution physically.

\newsec{Constraints on initial conditions}

A condition to solve the horizon problem 
(in the Einstein frame, thus the tildes) is 
\ref\rtres{Y.Hu, M. Turner and E. Weinberg, 
{\it Phys. Rev.} {\bf D}49, 3830 (1994)}
\eqn\hor {d_{{\rm HOR}}(\tilde t_f)={\tilde a}(\tilde t_f)
\int_{\tilde t_i}^{\tilde t_f}{d\tilde t'/{\tilde a}(\tilde t')}>{\tilde a}
(\tilde t_f)H_0^{-1}/{\tilde a}_0} where $H_0^{-1}$ is 
the size of the observed Universe ($H_0^{-1}\sim 10^{28}$cm),  
  $\tilde t_f$ is the time by which the horizon problem is 
  solved and $\tilde t_i$ the time when inflation began. 
In order to solve the horizon problem, this condition, equation \hor\ , must be satisfied in the Einstein frame as well as in the string frame. We compute our calculations in the string frame. For us, $t_i$ and $t_f$ determine  
the time range when the pre-big-bang description we 
are using remains valid. 
There is little reason to expect $t_i$ to be anything 
else other than $-\infty$ 
: one of the conceptual beauties of the pre big bang 
scenario is that the 
inital state is empty flat space, a very perturbative 
string vacuum. As for 
$t_f$, it should be of order $-t_{\rm Planck}$, 
we will estimate it carefully 
below. 
 Obviously, letting $t_i\to - \infty$ we see from 
  \hor\   that the horizon 
problem is solved both for $k=0$.  
 Still, it is of some 
interest to ask   how long did the universe have to 
behave stringly before the 
big bang in order for it to come out free of flatness 
and horizon problems from 
the high curvature epoch (around $t=0$). 
 
Let us now compute the amount of expansion required 
to solve the horizon 
problem, which is given  by the ratio  \rthree: 
\eqn\factordeinflacion{Z={H(t_f)a(t_f)\over H(t_i)a(t_i)}}
Experimentally (or rather, observationally), we need
\eqn\sesenta{Z>e^{60}}
in order to solve the horizon problem for our big universe.

 In the string frame the amount of inflation for a 
 flat space using \soltres\ turns out to be
\eqn\zeta {Z=\left ({(-t_{i})\over (-t_{f})} \right )^{3/2}}
 using \soltrese\ it is also true that
\eqn\zetacola{Z=\left ( {e^{-\phi (t_i)} \over 
e^{-\phi (t_f)}}\right )^{1/2}}
From   \sesenta\ and \zeta\   , it follows that
\eqn\implicacion{t_i<e^{40}t_f}
and
\eqn\implicaciono{ e^{\phi (t_f)}>e^{120}e^{\phi (t_i)}}

Since we are dealing with low-energy effective action we 
have two constraints on the solutions obtained from for 
the time when inflation ends $t_f$. Since our  effective 
action stops being valid when  gravity becomes 
strongly coupled, we expect the pre big bang inflationary 
epoch to be over by 
the time $t_f$ when 
 \eqn\restriccionuno{e^{\phi (t_f)}\leq 1}
Similarly, the same effective action remains valid only  
while the curvature is not too big:
\eqn\restricciondos {H^{-1}(t_f )\sim (-t_f)\geq l_{st}}
When $k=0$, these two requirements coincide, and the amount 
of inflation is thus
\eqn\zetacero {Z=\left ( {-t_i\over l_{{\rm st}}}\right )^{3/2} }
which means that $t_i<-10^{17}l_{\rm st}$  in order to 
solve the horizon problem. Note that as $t_i<t<t_f$ it 
is true that $e^{\phi (t_f)}<1$

From the solutions for $k=-1$, equations 
\soluno\ and \soldos, we notice that in both 
cases $e^{\phi (t_i)}$ can be very large, but 
our equations remain valid only as long as  
$e^{\phi(t_i)}\geq 1$ which implies $t_i>-10^{-9}\ell_{\rm st}$ for 
the first solution \soluno\ ridiculously small and even more 
ridiculously $t_i >-10^{-27}\ell_{\rm st}$ for the second one \soldos. 

The calculation of the amount of inflation $Z$ 
for negative curvature is difficult because there 
is no  accelerated  solution, that is why we would rather
 not  use a naive one. In any case the interval of time in 
 which our solutions  is reliable, $t_i<t<t_f$, is 
 extremely small since $G_N$ is decreasing with time, 
 which means that the time by which the universe starts 
 to expand $t_i$
has to be close to the big bang $t=0$. Because this is 
not an accelerated solution we do not have enough time to 
solve the horizon problem. 
It would seem that  nothing works out for  negative 
curvature, but this is actually good news. 
It is clear that, for some time, we can solve the 
horizon problem if we start with a flat region. 
Now,  a region with negative/positive curvature 
 between an infinitely big flat space cannot inflate 
 and get into what our observable universe is today. 
 In other words, it seems that our radiation dominated 
 universe comes from a flat  rather than a curved region.
 Of course we would rather not fall into an anthropic explanation,
   but initial conditions are rather hard to explain, unless we do
   it {\sl a posteriori}.
\newsec{Conclusions}
As \zetacero\ shows, there is a restriction on the beginning 
of the pre-big bang inflation in order to solve 
the horizon problem. For negative curvature, however, it is not
 posible to solve it. We attribute this to two reasons. 
 First, we do not get an accelerated phase, and as  
 standard cosmology says, we need an accelerated phase in 
 order to solve the horizon problem. Secondly,  as we go 
 back in time, the dilaton  increases (opposite to that for 
 flat space) and our tree-level low-energy effective action 
 \action\ breaks down.

Pre-big-bang cosmology requires a flat space 
\ref\rultima{G. Veneziano, talk given at 
International School of Subnuclear Physics
35th Course: Highlights: 50 Years Later, Erice, Italy, 
26 Aug - 4 Sep 1997.  hep-th/9802057} in order to solve 
succesfully the horizon problem, which is a much more 
appealling initial state than the hot and highly-curved space 
postulated by standard (inflationary) cosmology.

\listrefs
\bye